\documentclass[preprint,12pt]{elsarticle}

\usepackage{graphicx}
\usepackage{tikz}
\usepackage{color}

\usepackage{amsmath}
\usepackage{mathrsfs}
\usepackage{bm}

\newcounter{bla}

\journal{arXiv}

\begin{document}

\begin{frontmatter}



\title{A modified FDTD algorithm for processing ultra-wide-band response}


\author[a,b]{Huicheng Guo}
\author[a,b]{Henglei Du}
\author[a,b]{Chengpu Liu\corref{author}}

\cortext[author] {Corresponding author.\\\textit{E-mail address:} chpliu@siom.ac.cn (Chengpu Liu)}
\address[a]{State Key Laboratory of High Field Laser Physics and CAS Center for Excellence in Ultra-intense Laser Science, Shanghai Institute of Optics and Fine Mechanics, Chinese Academy of Sciences, Shanghai 201800, China}
\address[b]{Center of Materials Science and Optoelectronics Engineering, University of Chinese Academy of Sciences, Beijing 100049, China}

\begin{abstract}
  Finite-difference time-domain (FDTD) is an effective algorithm for resolving Maxwell equations directly in time domain. Although FDTD has obtained sufficient development, there still exists some improvement space for it, such as ultra-wide-band response and frequency-dependent nonlinearity. In order to resolve these troubles, a modified version of FDTD called complex-field frequency-decomposition (CFFD) FDTD method is introduced, in which the complex-field is adopted to eliminate pseudo-frequency components when computing nonlinearity and the frequency-decomposition is adopted to transform an ultra-wide-band response into a series of narrow-band responses when computing the interaction of ultra-short pulse with matters. Its successful applications in several typical situations and comparison with other methods sufficiently verify the uniqueness and superiority in processing ultra-wide-band response and frequency-dependent nonlinearity.

\end{abstract}


\end{frontmatter}




\begin{small}





\end{small}

\section{Introduction}
\label{sec:Introduction}
	Since the electromagnetic theory was established by Maxwell in 1873\cite{Classical_electrodynamics_by_Jackson}, Max\-well equations have been applied in a wide range of fields, such as communications, radar techniques, microwave circuits, laser physics, etc. The crucial using this theory to process engineering problems is to accurately resolve these equations at diverse boundary and initial conditions, while unfortunately only a few situations with symmetric structures could be calculated analytically. Instead, most realistic questions have to be approximately computed by numerical approaches. Meanwhile, with computer techniques emerging and progressing, many novel and valid algorithms that could effectively calculate electromagnetic phenomena have been proposed, e.g. finite element method (FEM)\cite{FEM_Jin}, method of moments (MoM)\cite{MoM_Gibson}, and finite-difference time-domain (FDTD)\cite{FDTD_1966_Yee}. Among these methods, due to the characteristic of its direct execution in time domain, FDTD naturally possesses distinct advantages to deal with wide-band spectral response from FEM and MoM methods.

	Since K. S. Yee proposed the FDTD method in 1966\cite{FDTD_1966_Yee}, it has obtained sufficient development\cite{FDTD_Taflove}, especially in aspects of Absorbing Boundary Conditions (ABCs)\cite{ABC_Mur}, Total Field and Scattering Field (TF/SF) conditions\cite{Total_scatering_field_Umashankar,Near_to_far_tranformation_Taflove} , Near-to-Far Field Transformation\cite{Near_to_far_tranformation_Taflove}, Piecewise-Linear Recursive Convolution (PLRC)\cite{PLRC_Kelley}, Auxiliary Differential Equation (ADE)\cite{ADE_Okoniewski},

 and so on. There yet exists some development space on elimination of  pseudo-frequency components when computing nonlinearity\cite{FDTD_nonlinear_shortage} and ultra-broad-band dispersion response that cannot be described by ideal models, like Lorentz, Drude, and Debye models\cite{FDTD_Taflove}. Hence, an efficient approach for resolving this puzzle is extremely necessary for those who are occupied in the interaction of ultra-short and ultra-strong laser with matters\cite{Femtosecond_laser_pulses}, which cannot be calculated accurately by traditional FDTD.

	For this, the complex-field frequency-decomposition (CFFD) method, a modified FDTD algorithm, is proposed, which includes two important steps---the complex-field step and the frequency-decomposition step. Fre\-quency-decompo\-sition step is used to transform an ultra-broad-band response that is beyond the ability of traditional FDTD methods, into a series of independent ideal responses that are convenient to be calculated. Simultaneously, complex-field step is used to eliminate the pseudo-frequency components that cannot be removed by traditional FDTD methods. Of course, if the response is broad-band but it is linear, one can choose complex-field step only to calculate this process, however the two steps must be adopted simultaneously when calculating wide-band nonlinear response. In addition, the CFFD-FDTD still involves two sub-approaches when calculating nonlinear problems, including perturbative CFFD and synchronously filtering CFFD. The former is suitable for computing perturbative nonlinear problems, while the latter is suitable for non-perturbative nonlinearities, although it is a little more complicated than perturbative CFFD method.

	What's more, several typical instances of CFFD used in ultra-wide-band linear dispersion and nonlinear second-harmonic generation (SHG) are presented to illuminate the advantages of CFFD method than traditional FDTD methods or the others. For linear situation, CFFD is used to simulate a Lorentz response that can be computed by ADE method, and an empirical response that cannot be computed by traditional FDTD methods, which sufficiently indicates the consistency with traditional method and the advantage of CFFD method, respectively. For nonlinear situation, we simulate the SHG processes under different conditions by adopting different methods. The CFFD can effectively eliminate the pseudo-frequency components, and the simulation of growth rate of SHG under the phase-matching conditions illustrates CFFD method has the advantage that it can accurately compute frequency-dependent nonlinearity in an ultra-wide-band response. The calculation of coherent length of SHG under phase-mismatching condition illustrates that CFFD method can accurately simulate the increasing and decreasing processes of electric amplitude as predicted by nonlinear theory.

	For more clearly illuminating the definite theory and application of CFFD method, the theoretical background of CFFD algorithm is provided in Section~\ref{sec:theoretical background}, which mainly includes decomposed Maxwell equations, approximations of polarization under linear and nonlinear conditions, single-carrier frequency approximation, perturbative nonlinear approximation and synchronously filtering method. In section~\ref{sec:Applications of CFFD algorithm}, some representative applications of CFFD have been successfully presented logically, and the validity and advantage of CFFD are confirmed practically. Finally, the main conclusion is drawn in Section~\ref{sec:conclusion}.

\section{Theoretical background}
\label{sec:theoretical background}
\subsection{Decomposition of Maxwell equations}

	It is well known that the dynamics of an electromagnetic wave in media is governed by Maxwell equations\cite{Classical_electrodynamics_by_Jackson},
	\begin{equation}\label{eq:Maxwell Equations}
		\begin{cases} 
			\nabla \times \bm H(t) = \dfrac{\partial \bm D(t)}{\partial t} + \bm J(t), \phantom{\bigg(} \\
			\nabla \times \bm E(t) = -\dfrac{\partial \bm B(t)}{\partial t} - \bm J_m(t), \phantom{\bigg(} \\
			\nabla \cdot \bm D(t) = \rho(t), \\
			\nabla \cdot \bm B(t) = \rho_m(t).
		\end{cases}
	\end{equation}
	Here, one point needs to note that free magneton $\rho_m$ and conductive magnetic current $\bm J_m$ are virtually added not only just  for the symmetry of equations, but also for the convenience of implementation, although the two quantities do not exist in practice. Hence, we can set them to zero finally so that they do not influence all calculations.  Eqs.\eqref{eq:Maxwell Equations} is easily transformed into one in frequency domain by Fourier transformation $\mathscr{F}\{\cdot\}$ as
	\begin{equation}\label{eq:Maxwell Equations in frequency domain}
		\begin{cases} 
			\nabla \times \bm H(\omega) =  i\omega \bm D(\omega) + \bm J(\omega), \phantom{\bigg(} \\
			\nabla \times \bm E(\omega) = -i\omega \bm B(\omega) - \bm J_m(\omega), \phantom{\bigg(} \\
			\nabla \cdot \bm D(\omega) = \rho(\omega), \\
			\nabla \cdot \bm B(\omega) = \rho_m(\omega).
		\end{cases}
	\end{equation}
	Electronic placement vector $\bm D(\omega) = \varepsilon_0 \varepsilon_r(\omega) \bm E(\omega)$, and permittivity $\varepsilon_r$ is a function of $\omega$. Generally,  $\bm D(\omega)$ is expressed via Taylor series as\cite{Nonlinear_Optics_boyd},
	\begin{equation}\label{eq:pertubative nonlinearity}
		\bm D(\omega) = \varepsilon_0 \bm E(\omega) + \varepsilon_0 \chi^{(1)}(\omega) \bm E(\omega) + \varepsilon_0 \chi^{(2)}(\omega;\omega_1,\omega_2) \bm E(\omega_1) \bm E(\omega_2) + \cdots.
	\end{equation}
The tensor $\chi^{(1)}(\omega)$ represents the linear response of media to an external field. If the response is limited within a relatively narrow frequency band, it can be formulated by one of three ideal dispersive models or their linear superposition. However, nonlinear tensors\cite{Nonlinear_Optics_boyd} $\chi^{(2)}(\omega;\omega_1,\omega_2), \cdots, \chi^{(m)}(\omega;\omega_1,\dots,\omega_m)$ are unable to be described in former formation, so they can not be directly used in the  conventional FDTD framework, except some special approximations such as Born-Oppenheimer approximation\cite{Nonlinear_approximation_Born_Oppenheimer}, which undoubtedly confines its application extent.

	An effective approach to resolve this difficulty is to cancel out the closely frequency-dependent characteristic of  $\varepsilon_r(\omega)$. To do this, we can decompose the whole frequency space into a series of sub-spaces by some window functions of  $f_{\pm i}(\omega)$ with constraint condition
	\begin{equation} \label{eq:decompositing frequency} 
		1 \equiv \sum_{i=1}^{\infty} \big[ f_{-i}(\omega) + f_{i}(\omega) \big],\quad \omega \in (-\infty,+\infty).
	\end{equation}
	For simplicity, the functions $f_{\pm i}(\omega)$ are chosen as rectangular functions and symmetric about zero frequency point, namely
	\begin{equation}
		f_{\pm i}(\omega) \equiv \mathrm{rect}(\frac{\omega \pm \omega_i}{L_i}) = \begin{cases}
		1,&|\omega \pm \omega_i| < L_i, \\
		0,&|\omega \pm \omega_i| \geq L_i. \end{cases}				
	\end{equation}
	Here, $\omega_{i}$ and $L_{i}$ represent center point and width of the $i$'th window function, and are defined as 
	\begin{equation}
		 \begin{cases} 
	\omega_{i+1}-\omega_{i} = L_{i+1} + L_{i}, \\ 
	\omega_{1} = L_1, \\ 
	\omega_{i},L_{i} > 0 ,\quad \forall ~i.
	\end{cases} 
	\end{equation}
	Subsequently, multiplying both side of Eq.\eqref{eq:Maxwell Equations in frequency domain} by $f_{\pm i}(\omega)$ gets
	\begin{equation} \label{eq:decompositional Maxwell equations}
		\begin{cases}
			\nabla \times \bm H_i(\omega) = i\omega \bm D_i(\omega) + \bm J_i(\omega), \\
			\nabla \times \bm E_i(\omega) = -i\omega \bm B_i(\omega) - \bm J_{m,i}(\omega), \\
			\nabla \cdot \bm D_i(\omega) = \rho_i(\omega), \\
			\nabla \cdot \bm B_i(\omega) = \rho_{m,i}(\omega). \\
		\end{cases}
	\end{equation}
The corresponding constitutive relations are written as
	\begin{equation} \label{eq:decompositional D and B}
		\begin{cases}
			\bm D_i(\omega) = \varepsilon_0 \bm E_i(\omega) + \bm P_{i}(\omega), \\
			\bm B_i(\omega) = \mu_0 \bm H_i(\omega) + \mu_0 \bm M_{i}(\omega). \\
		\end{cases}
	\end{equation}
	The field components with subscript ``$i$" have similar forms. For example, the electric field vector $\bm E_i(\omega)$  takes the form of $ \bm E_i(\omega) \equiv \bm E(\omega)f_i(\omega)$,  and electronic density scalar $\rho(\omega)$  of $\rho_i(\omega) \equiv \rho(\omega)f_i(\omega)$, and so on. Whereafter, taking inverse Fourier transformation of Eqs.\eqref{eq:decompositional Maxwell equations} and \eqref{eq:decompositional D and B} into time domain, obtains decomposed Maxwell equations
	\begin{equation}\label{eq:decompositional Maxwell equations in time}
		\begin{cases}
			\nabla \times \bm H_i(t) = \dfrac{\partial}{\partial t} \bm D_i(t) + \bm J_i(t), \phantom{\bigg(}\\
			\nabla \times \bm E_i(t) = -\dfrac{\partial}{\partial t} \bm B_i(t) - \bm J_{m,i}(t), \phantom{\bigg(}\\
			\nabla \cdot \bm D_i(t) = \rho_i(t), \\
			\nabla \cdot \bm B_i(t) = \rho_{m,i}(t),
		\end{cases}
	\end{equation}
and decomposed constitutive relations
	\begin{equation}\label{eq:decompositional D and B in time}
		\begin{cases}
			\bm D_i(t) = \varepsilon_0 \bm E_i(t) + \bm P_{i}(t), \\
			\bm B_i(t) = \mu_0 \bm H_i(t) + \mu_0 \bm M_{i}(t).
		\end{cases}
	\end{equation}
Eqs.~\eqref{eq:decompositional Maxwell equations in time} and \eqref{eq:decompositional D and B in time} describe evolution of field components only within the $f_{i}$ window, which are called CFFD equations. The biggest difference between Eqs.~\eqref{eq:decompositional Maxwell equations in time} and Eqs.~\eqref{eq:Maxwell Equations} is that the field components out of  the $f_{i}$ window identically equal to zero, while those within the $f_{i}$ window are completely same with the initial fields. In addition, if non-magnetic medium is considered, the quantities $\bm M_{i}$, $\bm J_{m,i}$ and $\rho_{m,i}$ vanish completely. Furthermore, according to nonlinear theory, electronic polarization could be expressed as $\bm P_i(\omega) = \bm P_{L,i}(\omega) + \sum \limits_{m=2}^{\infty} \bm P_{NL,i}^{(m)}(\omega)$, so the next crucial works are to analyze these electronic polarizations $ \bm P_{L,i}(\omega) $ and $\bm P_{NL,i}^{(m)}(\omega)$. However, if the medium is magnetic, the quantities $\bm M_{i}$, $\bm J_{m,i}$ and $\rho_{m,i}$ can be analyzed like the process analyzing electronic polarizations.

\subsection{Linear polarization approximation}
\label{sec:Linear polarization approximation}
	First, it is convenient to analyze linear situations where nonlinearity is neglected. In an ultra-wide band, it could hardly accurately describe dispersion of dielectrics through superposition of several ideal models. Furthermore, if resonant frequency is too large, the efficiency and accuracy of computation will decrease significantly. Based upon this, frequency-decomposition method splits an ultra-wide band into a series of narrow ones where their dispersion can be represented by a few responses convenient to be calculated. For simplicity, we analyze an isotropic dielectric, and the tensor $\chi^{(1)}_i$ degrades into a scalar, so the linear polarization is formulated as
	\begin{equation} \label{eq:linear polarization}
		\bm P_{L,i}(\omega) = \varepsilon_0 [\chi^{(1)}_i(\omega_i) + \delta_i(\omega)] \bm E_i(\omega),
	\end{equation}
	where $\chi^{(1)}_i(\omega_i)$ represents the central response of window function $f_i(\omega)$, and $\delta_i(\omega)$ is a function describing the deviation of $\chi^{(1)}_i(\omega_i)$ from the realistic response within the interval $[\omega_i-L_i,\omega_i+L_i]$. If $\delta_i(\omega)$ is small enough in the interval, it can be approximated as zero. It means that the phase error introduced by ignoring $\delta_i(\omega)$ is also negligible. Therefore, $\bm P_{L,i}(\omega)$ can be approximated as
	\begin{equation} \label{eq:linear polarization constant approximation}
		\bm P_{L,i}(\omega) = \varepsilon_0 \chi^{(1)}_i(\omega_i) \bm E_i(\omega).
	\end{equation}
	Then, inverse transformation of Eq.~\eqref{eq:linear polarization constant approximation} back into time domain obtains
	\begin{equation} \label{eq:linear polarization constant approximation time domain}
		\bm P_{L,i}(t) = \varepsilon_0 \chi^{(1)}_i(\omega_i) \bm E_i(t).
	\end{equation}

	However, when $\delta_i(\omega)$ cannot be ignored, it can also be approximated by several typical models as
	\begin{equation} \label{eq:linear polarization classical approximation}
		\bm P_{L,i}(\omega) = \varepsilon_0 [\chi^{(1)}_i(\omega_i) + \tilde\delta_i(\omega)] \bm E_i(\omega),
	\end{equation}
	where $\tilde\delta_i(\omega)$ obeys
	\begin{equation} \label{eq:approximate delta function}
		\tilde\delta_i(\omega) = \dfrac{\sum \limits_k p_k (i\omega)^k}{\sum \limits_l q_l (i\omega)^l},
	\end{equation}
	in which $p_k$ and $q_l$ are the expanded coefficients. It is noted that error outside this interval ought not to matter, because $\bm E_i(\omega)$ identically vanishes there. Similarly, inverse transformation of Eq.~\eqref{eq:linear polarization classical approximation} into time domain obtains
	\begin{equation} \label{eq:linear polarization classical approximation time domain}
		\bm P_{L,i}(t) = \varepsilon_0 [\chi^{(1)}_i(\omega_i) \delta(t) + \tilde\delta_i(t)] * \bm E_i(t),
	\end{equation}
	where $\delta(t)$ is Dirac function and $\tilde\delta_i(t) = \mathscr{F}^{-1}\{ \tilde\delta_i(\omega) \}$ represents ideal response that is convenient to be calculated by traditional FDTD method.

\subsection{Nonlinear polarization approximation}
	When nonlinearity is non-negligible and evidently depends on frequency in an ultra-wide band, CFFD method also could decompose it into a series of narrow bands where their nonlinearity can be represented by a few conveniently calculated responses. Analyzing isotropic system for convenience, $\chi^{(m)}(\omega;\omega_{1},\dots,\omega_{m})$ is a scalar and the corresponding nonlinear polarization is expressed as
	\begin{equation} \label{eq:nonlinear polarization}
		\tilde {\bm P}_{i}^{(m)}(\omega) = \varepsilon_0 [\chi^{(m)} + \delta_i^{(m)}(\omega;\omega_1,\dots,\omega_m)] \tilde {\bm E}_{1i}(\omega_1) \cdots \tilde {\bm E}_{mi}(\omega_m),
	\end{equation}
	where symbol `\raisebox{-0.4em}[0.1em][0.2em]{$\tilde{\phantom{a}}$}' implies it is a complex quantity. $\chi^{(m)} = \chi^{(m)}(\omega_i;\omega_{1_i},\dots,\omega_{m_i})$ is the central response of $m$-dimensional window function $f_{i_1}(\omega_1) \cdots f_{i_m}(\omega_m)$. 
$\delta_i^{(m)}(\omega;\omega_1,\dots,\omega_m)$, as done in the linear approximation in Section \ref{sec:Linear polarization approximation}, is also a scalar function describing the deviation of $\chi^{(m)}$ from the realistic response within  $m$-dimensional interval $[\omega_{i_1}-L_{i_1},\omega_{i_1}+L_{i_1}] \times \cdots \times [\omega_{i_m}-L_{i_m},\omega_{i_m}+L_{i_m}]$. 

	In the similar manner mentioned in Section \ref{sec:Linear polarization approximation}, if $\delta_i^{(m)}$ is small in the interval, it can be approximated as zero, which means phase error introduced by ignoring $\delta_i^{(m)}$ is negligible. Therefore, $\bm P_{i}^{(m)}(\omega)$ can be approximated as
	\begin{equation} \label{eq:nonlinear polarization constant approximation}
		\tilde {\bm P}_{i}^{(m)}(\omega) = \varepsilon_0 \chi^{(m)} \tilde {\bm E}_{1i}(\omega_1) \cdots \tilde {\bm E}_{mi}(\omega_m),
	\end{equation}
	and then, inverse transformation of Eq.\eqref{eq:nonlinear polarization constant approximation} into time domain by $m$-dimen\-sional Fourier transformation obtains
	\begin{equation} \label{eq:nonlinear polarization constant approximation time domain}
		\tilde{\bm P}_{i}^{(m)}(t) = \varepsilon_0 \chi^{(m)} \tilde{\bm E}_{1i}(t) \cdots \tilde{\bm E}_{mi}(t).
	\end{equation}

	 When $\tilde{\delta}_i^{(m)}(\omega;\omega_1,\dots,\omega_m)$ is important, it can be approximated by several simple dispersive models. Therefore, expression \eqref{eq:nonlinear polarization} can be locally expanded, according to the analysis theory of mathematics, by several linear dispersion functions as
	\begin{equation} \label{eq:nonlinear polarization classical approximation}
		\tilde{\bm P}_{i}^{(m)}(\omega) = \varepsilon_0 [\chi^{(m)} + \tilde\delta_{i_1}(\omega_1) \cdots \tilde\delta_{i_m}(\omega_m)] \tilde {\bm E}_{1i}(\omega_1) \cdots \tilde {\bm E}_{mi}(\omega_m),
	\end{equation}
	where $\tilde\delta_{i_j}(\omega_j)$ obeys
	\begin{equation} \label{eq:nonlinear approximate delta function}
		\tilde\delta_{i_j}(\omega_j) = \dfrac{\sum \limits_k p_k (i\omega_j)^k}{\sum \limits_l q_l (i\omega_j)^l}.
	\end{equation}
	It means that $\tilde{\delta}_i^{(m)}(\omega;\omega_1,\dots,\omega_m)$ can also be locally expanded as multiplications of several independent linear responses, in which $p_k$ and $q_l$ are the corresponding expanded coefficients. Like the former, errors outside the $m$-dimensional interval ought not to matter. Finally, inverse transformation of Eq.~\eqref{eq:nonlinear polarization classical approximation} into time domain obtains
	\begin{equation} \label{eq:nonlinear polarization classical approximation time domain}
	\begin{split}
		\hspace{-0.1cm} \tilde{\bm P}_{i}^{(m)}(t) &= \varepsilon_0 [\chi^{(m)} \delta(t) + \tilde\delta_{i_1}(t) \cdots \tilde\delta_{i_m}(t)] * [\tilde{\bm E}_{i_1}(t) \cdots \tilde{\bm E}_{i_m}(t)] \\
		&= \varepsilon_0 \chi^{(m)} \tilde{\bm E}_{i_1}(t) \cdots \tilde{\bm E}_{i_m}(t) + [\tilde\delta_{i_1}(t) * \tilde{\bm E}_{i_1}(t)] \cdots [\tilde\delta_{i_m}(t) * \tilde{\bm E}_{i_m}(t)],
	\end{split}
	\end{equation}
	where $\delta(t)$ is Dirac function and $\tilde\delta_{i_j}(t) = \mathscr{F}^{-1}\{ \tilde\delta_{i_j}(\omega_j) \}$ is the response convenient to be calculated by traditional FDTD method.

\subsection{Synchronously filtering method}
\label{subsec:Synchronously filtering method}
	According to the formulas of \eqref{eq:nonlinear polarization constant approximation time domain} and \eqref{eq:nonlinear polarization classical approximation time domain}, the final frequency range of nonlinear polarization $\tilde{\bm P}_{i}^{(m)}(t)$ overflow the $i$'th frequency interval. In order to filter out the overflowed components, a synchronously filtering method is necessary to be adopted. The types of filtering schemes are introduced respectively in the following.

\subsubsection{Spatial filtering}
	As for nonlinear approximation, when linear dispersion is ignored, time frequency  $\omega$ is uniquely related to spatial wavenumber $k$ via $k  = \omega/v = \omega \sqrt{\varepsilon \mu}$. Therefore, filtering for formulations \eqref{eq:nonlinear polarization constant approximation time domain} and \eqref{eq:nonlinear polarization classical approximation time domain} in frequency space can be transformed into spatial wavenumber space, via transformation of temporal filtering function $f_i(\omega)$ into spatial formation,
	\begin{equation}
		f_i(\omega) = \begin{cases} 1,\quad |\omega-\omega_i| < L_i, \\
		0,\quad others,	\end{cases} = \begin{cases} 1,\quad |k-k_i| < L_i/v, \\
		0,\quad others.	\end{cases}
	\end{equation}
	Then, we define the spatial filtering function $F_i(\bm k)$ as\cite{SF-FDTD1,SF-FDTD2,SF-FDTD3}
	\begin{equation} \label{eq:spatial filtering function}
		F_i(\bm k) = \begin{cases} 1,\quad \big||\bm k|-k_i\big| < L_i/|\bm v|, \\
		0,\quad others,	\end{cases}
	\end{equation}
	where $\bm k$ is the wave vector, $\bm v$ is corresponding phase velocity, and the filtering range is a spherical shell. By multiplying the spatial filtering function with $\tilde{\bm P}_{i}^{(m)}(\bm k,t) = \mathscr{F}_{3D}\{\tilde{\bm P}_{i}^{(m)}(\bm r,t)\}$ and taking inverse Fourier transformation, one can obtain the nonlinear polarization in space domain as
	\begin{equation} \label{eq:spatial filtering polarization}
		\tilde{\bm P}_{i}^{(m)}(\bm r,t) = \varepsilon_0 \chi^{(m)} \mathscr{F}_{3D}^{-1} \{ \mathscr{F}_{3D}\{ \tilde{\bm E}_{1i}(\bm r,t) \cdots \tilde{\bm E}_{mi}(\bm r,t) \} F_i(\bm k) \}.
	\end{equation}

\subsubsection{Temporal filtering}
	However, if linear dispersion is included, the above spatial filtering method fails to process nonlinearity, because time frequency cannot be uniquely correlated to a fixed spatial wavenumber. The temporal filtering method is thus introduced, which constructs a virtual dispersive function $\varepsilon_{r,i}(\omega)$ with characteristics of extremely large loss outside the $i$'th frequency interval but without any loss inside this interval. Then, addition of the virtual function $\varepsilon_{r,i}(\omega)$ to linear polarization expression \eqref{eq:linear polarization classical approximation} obtains modified formation
	\begin{equation} \label{eq:time filtering linear polarization classical approximation}
		\bm P_{L,i}(\omega) = \varepsilon_0 [\chi^{(1)}_i(\omega_i) + \tilde\delta_i(\omega) + \varepsilon_{r,i}(\omega)] \bm E_i(\omega).
	\end{equation}
Here, $\varepsilon_{r,i}(\omega)$ can be constructed by some ideal models. Now, inverse transformation of Eq.~\eqref{eq:time filtering linear polarization classical approximation} back into time domain obtains
	\begin{equation} \label{eq:time filtering linear polarization classical approximation time domain}
		\bm P_{L,i}(t) = \varepsilon_0 [\chi^{(1)}_i(\omega_i) \delta(t) + \tilde\delta_i(t) + \varepsilon_{r,i}(t)] * \bm E_i(t).
	\end{equation}
	Subsequently, we substitute Eq.~\eqref{eq:time filtering linear polarization classical approximation time domain} into Eq.~\eqref{eq:decompositional D and B in time} and utilize the routine PLRC method\cite{FDTD_Taflove} to calculate this convolution.

\subsection{Single carrier frequency approximation}
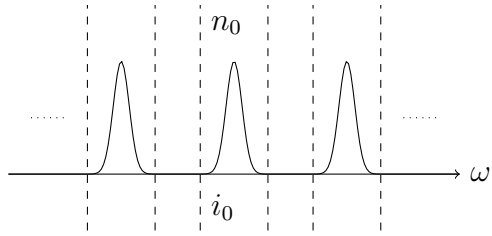
\begin{figure}[htbp]
		\centering
		\begin{tikzpicture}[scale = 1.5]
		\draw[->] (-2,0) - - (2,0) node[right] {$\omega$};
		\draw[domain = -2:2,samples = 200] plot(\x,{exp(-(\x/0.1)^2) + exp(-((\x-1)/0.1)^2) + exp(-((\x+1)/0.1)^2)} );
		\draw[dashed] (-1.3,-0.5) - - (-1.3,1.5);
		\draw[dashed] (-0.7,-0.5) - - (-0.7,1.5);
		\draw[dashed] (-0.3,-0.5) node[above right] {$i_0$} - - (-0.3,1.5) node[below right] {$n_0$};
		\draw[dashed] (0.3,-0.5) - - (0.3,1.5);
		\draw[dashed] (0.7,-0.5) - - (0.7,1.5);
		\draw[dashed] (1.3,-0.5) - - (1.3,1.5);

		\draw[dotted] (-1.8,0.5) - - (-1.5,0.5);
		\draw[dotted] (1.5,0.5) - - (1.8,0.5);
		\end{tikzpicture}
		\caption{Single carrier frequency approximation}
		\label{fig:SCFA}				
	\end{figure}

	Single carrier frequency approximation (SCFA) is from the work of Refs.~\cite{SCFA_Ref1,SCFA_Ref2,SCFA_Ref3}. Here, we present a detailed derivation of it. Given that an input pulse includes $N$ carrier frequencies that are clearly separated from each other, as shown in Fig.~\ref{fig:SCFA},
	the electric field within the $i_0$'th frequency interval can be approximately written as
	\begin{equation} \label{eq:SCFA formulation}
		\begin{split}
			\tilde{\bm E}_{i_0}(t) &= \tilde{\bm E}_{n_0,i_0}(t) + \sum_{n \neq n_0} \tilde{\bm E}_{n,i_0}(t) \approx \tilde{\bm E}_{n_0,i_0}(t) \\
&\approx \tilde{\bm E}_{n_0,i_0}(t) + \sum_{i \neq i_0} \tilde{\bm E}_{n_0,i}(t) = \sum_i \tilde{\bm E}_{n_0,i}(t) \\
&= \tilde{\bm E}_{n_0}(t).
		\end{split}
	\end{equation}
Here, $\tilde{\bm E}_{i}(t)$ is the total complex field within $i$'th frequency interval, $\tilde{\bm E}_{n}(t)$ is the complex field with carrier frequency $\omega_{n}$, and $\tilde{\bm E}_{n,i}(t)$ is the complex field with the $n$'th carrier frequency $\omega_{n}$ and within the $i$'th frequency interval. According to Eq.~\eqref{eq:SCFA formulation}, the electric field $\tilde{\bm E}_{i_0}(t)$ can be finally approximated as $\tilde{\bm E}_{n_0}(t)$, so the total field can be decomposed as a sequence of sub-fields $\tilde{\bm E}_{n}(t)$ to be processed independently under condition of single-carrier frequency approximation.

\subsection{Perturbative nonlinearity}
	If nonlinearity is perturbative and the band of excited pulse is narrow according to Eq.~\eqref{eq:SCFA formulation}, nonlinear polarization can be approximately described as
	\begin{equation} \label{eq:SCFA polarization time domain}
		\tilde{\bm P}_{n}^{(m)}(t) = \varepsilon_0 \chi^{(m)} \tilde{\bm E}_{n_1}(t) \cdots \tilde{\bm E}_{n_m}(t),
	\end{equation}
	which is very similar to Eq.~\eqref{eq:nonlinear polarization constant approximation time domain}. The difference is that Eq.~\eqref{eq:SCFA polarization time domain} is decomposed by carrier frequency $\omega_{n}$, while Eq.~\eqref{eq:nonlinear polarization constant approximation time domain} is by frequency interval. This distinction will become more obvious under non-perturbative and broad-band conditions. The curl equations and constitutive equations based on Eq.~\eqref{eq:SCFA polarization time domain} become
	\begin{equation}\label{eq:SCFA Maxwell equations in time}
		\begin{cases}
			\nabla \times \bm H_n(t) = \dfrac{\partial}{\partial t} \bm D_n(t) + \bm J_n(t), \phantom{\bigg(}\\
			\nabla \times \bm E_n(t) = -\dfrac{\partial}{\partial t} \bm B_n(t) - \bm J_{m,n}(t), \phantom{\bigg(}\\
			\bm D_n(t) = \varepsilon_0 \bm E_n(t) + \bm P_{n}(t), \\
			\bm B_n(t) = \mu_0 \bm H_n(t) + \mu_0 \bm M_{n}(t).
		\end{cases}
	\end{equation}

\section{Applications of CFFD algorithm}
\label{sec:Applications of CFFD algorithm}
	For illuminating the advantages of CFFD algorithm proposed in section~\ref{sec:theoretical background}, several typical examples are presented in order. First, we simulated broad-band linear response under Lorentz and empirical dispersive functions to present its advantages of computing linear problems. Then, we simulate second-order nonlinearity from three aspects, including elimination of pseudo-frequency components, frequency-dependent nonlinearity, and mismatched phase conditions, which can illustrate the advantages of CFFD algorithm in processing nonlinear problems.

\subsection{Lorentz response}
\label{subsec:Lorentz response}
\begin{figure} [ht]
		\centering
		\includegraphics[scale=0.8]{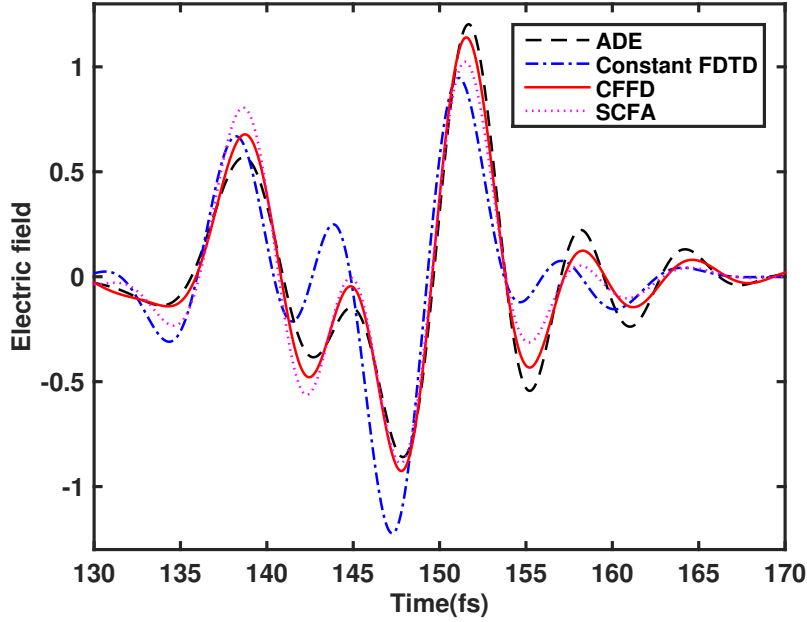}
		\caption{ \label{fig:Lorentz model} Electric fields after propagation of 18 $\mu$m by four different algorithms.} 
	\end{figure}

	In order to determine the accuracy of CFFD algorithm, we simulate the propagation of laser pulse within Lorentz dielectric by CFFD algorithm and compare the results with that given by ADE algorithm, as shown in Fig.~\ref{fig:Lorentz model}. For Lorentz response, it has been known that Maxwell equations can be accurately calculated by the ADE algorithm\cite{ADE_Okoniewski}. If our calculation based  on the CFFD algorithm is consistent with that of ADE, the validity can be confirmed. In addition, in Fig.~\ref{fig:Lorentz model} we also include the simulations based on the other two algorithms for further comparison, i.e. the constant FDTD and the SCFA. Here the constant FDTD indicates that the electric permittivity is a constant, which equals to the CFFD algorithm with only one frequency-decomposition interval. In contrast, the simulation implemented by the CFFD algorithm adopts five inhomogeneous frequency-decomposition intervals of $[0,70)$, $[70,80)$, $[80,145)$, $[145,155)$, and $[155,+\infty)$. The Lorentz response is given by\cite{ADE_Okoniewski} 
	\begin{equation}\label{eq:LorentzResponse}
		\chi_l(\omega) = \frac{(\varepsilon_s - \varepsilon_{\infty})\omega_0^2}{\omega_0^2 + 2j\nu_c\omega - \omega^2},
	\end{equation}
	so the relative permittivity is $\varepsilon(\omega) = \varepsilon_{\infty}+\chi_l(\omega)$. Parameters of $\omega_0 = 2\pi\nu_0$, $\nu_0 = 550$~THz, $\nu_c = 0$~THz, $\epsilon_s = 2$, $\epsilon_{\infty} = 1$ are chosen for simulation. In addition, the incident laser pulse with two carrier frequencies (two-color laser) is defined as
	\begin{equation}\label{eq:two-color pulse}
		\bm E_i(t) = \bm E_0 e^{(\frac{t-t_0}{T_w})^2} [ \cos(\omega_{i_1} t) + \cos(\omega_{i_2} t + \phi_0) ],
	\end{equation}
where $f_{i1}=75~\mathrm{THz}$, $f_{i2}=150~\mathrm{THz}$, $\omega_{i1}=2\pi f_{i1}$, $\omega_{i2}=2\pi f_{i2}$, $T_w=10~\mathrm{fs}$, $t_0=6 T_w$, and $\phi_0 = 0$.
	
	Now, by substituting Eqs.~\eqref{eq:LorentzResponse} and \eqref{eq:two-color pulse} into Eqs.~\eqref{eq:linear polarization constant approximation time domain}, and solving Eq.~\eqref{eq:decompositional Maxwell equations in time} via using the above four algorithms, we can investigate the field waveforms of the two-color laser after propagating certain distance (18~$\mu$m,$\sim$ 10 laser wavelengths here) indicated in Fig. ~\ref{fig:Lorentz model}. 

From Fig.~\ref{fig:Lorentz model}, one can see that the electric field obtained by constant FDTD (dash-dotted) obviously deviates from the accurate solution by ADE (dashed). In contrast, the result of CFFD (solid) is closer to that of ADE, as the consequence of usage of more frequency-decomposition intervals, which indicates that CFFD provides a more accurate solution than constant FDTD. The point to be noted is that the frequency-decomposition intervals adopted in CFFD are much coarser, and if a finer decomposition is chosen, the solution given by CFFD will and should be expected to be completely consistent with that by ADE. In addition, the SCFA result in Fig.~\ref{fig:Lorentz model} (dotted) obviously deviates from that by ADE, which is the consequence of phase errors introduced of SCFA due to the short pulse duration. If the pulse duration is long enough, the result of SCFA will approach to the accurate one. Therefore, the comparison among different algorithms shows that our proposed CFFD algorithm is superior to SCFA and constant FDTD, and comparable with the widely used ADE algorithm.

In fact, the above comparison can be extended to the other typical Drude or Debye dielectric response model, because the conventional ADE as well as PLRC and Z-transformation algorithms can also provide pretty accurate solutions to Maxwell equations. This comparison among these algorithms indicates that the CFFD algorithm can be successfully used for these models. However, if the dielectric response, like Cauchy's empirical model, can not be described by these typical models, these conventional algorithms fail and only our CFFD algorithm can still be used.

\subsection{Empirical response}
\label{subsec:Empirical response}
	The Cauchy-like empirical response model\cite{PMMA_Ref} formulating the relationship between refractive index $n$ and laser wavelength $\lambda$ is described as 
	\begin{equation}
		n^2(\lambda) = a_0 + a_1 \lambda^2 + a_2 \lambda^4 + a_3 \lambda^{-2} + a_4 \lambda^{-4} + a_5 \lambda^{-6} + a_6 \lambda^{-8},
	\end{equation}
	and the coefficients $a_i$ ($i=0,\cdots,6$) are taken from Ref.~\cite{PMMA_Ref}.

According to Fresnel theory\cite{PrinciplesOfOptics_Born}, analytical transmission and reflection coefficients for a pulse normally incident on this Cauchy-like dielectric from vacuum are written as
	\begin{equation} \label{eq:Fresnel formulations}
	\begin{split}
		|r(\nu)| &\equiv \big| \frac{E_{r0}(\nu)}{E_{i0}(\nu)} \big| = \frac{n(\nu)-1}{n(\nu)+1}, \\
		|t(\nu)| &\equiv \big| \frac{E_{t0}(\nu)}{E_{i0}(\nu)} \big| = \frac{2}{n(\nu)+1}.		 
	\end{split}
	\end{equation}
	where the terms on the right side of  ``$\equiv$'' in Eqs.~\eqref{eq:Fresnel formulations} are used for numerical calculation, while the terms on the right side of ``=" are used for analytical calculation of transmission and reflection coefficients. The consistency between the numerical and analytical results can directly confirm the validity and advantage of CFFD algorithm.

\begin{figure} [ht]
		\begin{center}
		\hspace*{-0.4cm}
		\begin{tabular}{c}
			\includegraphics[scale=0.65]{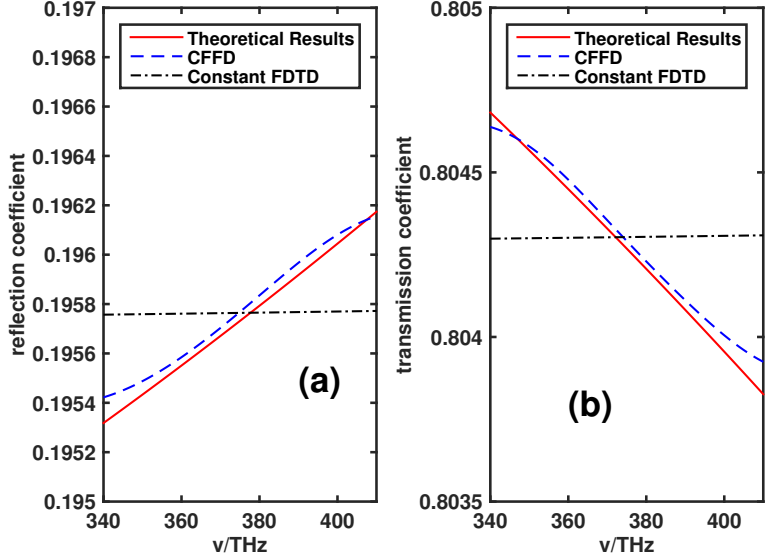}
		\end{tabular}
		\end{center}
		\caption{ \label{fig:Cauchy model}(a) Reflection coefficient; (b) Transmission coefficient.} 
	\end{figure}

	As for the numerical calculation, an ultra-short pulse with single carrier frequency is provided as
	\begin{equation}
		\bm E(t) = \bm E_0 e^{\frac{(t-t_0)^2}{T_w^2}} cos(2 \pi \nu_0 t),
	\end{equation}
	where $\nu_0 = 375$~THz and $T_w = 3.4$~fs. We decompose this pulse in frequency domain into nine frequency intervals of $[0,340)$, $[340,350)$, $[350,360)$, $\dots$, $[400,410)$, and $[410,+\infty)$~THz. The response centers corresponding to these intervals are selected at $335$~THz, $345$~THz, $355$~THz,$\dots$, $405$~THz, $415$~THz, respectively. The coefficients obtained by CFFD and the theoretical calculation are shown in Fig.~\ref{fig:Cauchy model}, where the numerical result by constant FDTD is also shown just for comparison.

	From Fig.~\ref{fig:Cauchy model}, we know that the result by CFFD is basically consistent with the accurate theoretical one. The results are more consistent within the frequency range from 360~THz to 400~THz, because here the frequency intervals are much finer than that outside these frequency ranges. Of course, if a finer frequency decomposition is adopted, the solution given by CFFD will and should be expected to be completely consistent with the theoretical result. In contrast, the result given by constant FDTD is almost invariant with frequency, which obviously deviates from the correct results provided by theoretical calculation and our CFFD algorithm.

\subsection{Elimination of pseudo-frequency components}
\label{subsec:Elimination of pseudo-frequency components}
	The CFFD algorithm is not only suitable for ultra-wide-band linear problems, but also for nonlinear problems. The biggest difference between nonlinear calculation and linear calculation, according to nonlinear theory, is the introduction of complex field instead of real field.	Here, the real field refers to an actual vibration, e.g.~$\sin(\omega t)$, with non-negative frequency $\omega \in [0,+\infty)$. In contrast, complex field refers to a virtual vibration, e.g.~$exp(-i\omega t)$, artificially constructed for conveniently taking Fourier analysis to completely characterize the whole field. Therefore, in practical applications, complex field is more widely used than real field.

	To highlight the defect of real field in FDTD calculation, the nonlinear response of polarization $\bm P(t)$ is considered. In nonlinear medium, different frequencies will mutually couple with each other, and thus a new  frequency component will be induced. If real field is adopted, it will inevitably make frequency simultaneously transferred upwards and downwards, so the pseudo-frequency components are induced. Their occurrence is in contradiction with theoretical analysis from the conventional nonlinear theory, and thus they are unphysical and merely artificial mathematical errors. It is very difficult to remove these additional frequency components if you still adopt real field. However, if you use complex field instead of real field, these pseudo-frequency components do not occur at all. This is the reason why we adopt complex field in CFFD algorithm.
	\begin{figure} [ht]
		\begin{center}
		\begin{tabular}{c}
			\hspace*{-0.65cm}
			\includegraphics[scale=0.55]{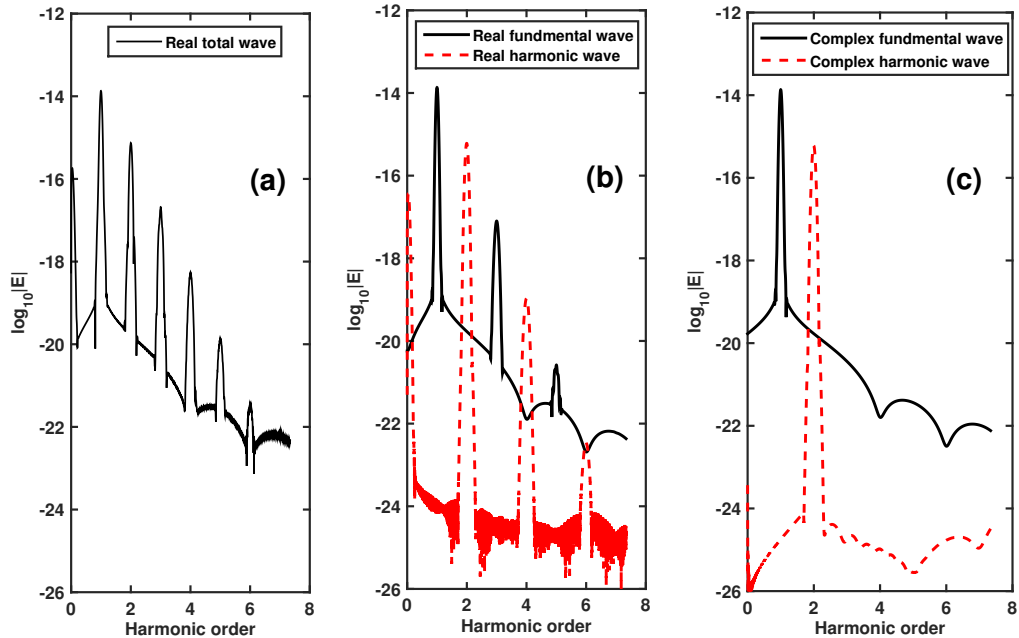}
		\end{tabular}
		\end{center}
		\caption{ \label{fig:CFFD necessity}(a) Real field without decomposition; (b) real field with decomposition; (c) complex field with decomposition.} 
	\end{figure}

	As an example, a second-order nonlinear process is investigated and only the second harmonic generation from the fundamental wave is focused. The respective simulations by complex field and real field are implemented and their results are compared. For the SHG process under real field without frequency decomposition, the real fundamental wave with $\omega$ first induces direct-current and second-order components due to $cos^2(\omega t) = [1+cos(2 \omega t)]/2$. Then, these two new frequencies are back-acted in the real fundamental field to further induce higher-order frequency components. This process is repeated ceaselessly and finally results in a train of pseudo-harmonics, as shown in Fig.~\ref{fig:CFFD necessity}(a). If now we adopt frequency decomposition and still use real field to describe the fundamental and second-order harmonic fields, these two fields are coupled with each other via nonlinearity. After many iterations of these two fields, the fundamental field becomes one only with a train of odd-order harmonics, and the second-order field becomes one with only even-order harmonics, shown in Fig.~\ref{fig:CFFD necessity}(b). These pseudo-frequency components still do not be removed. However, if we adopt complex field with frequency decomposition, the trouble of frequency upwards and downwards transferring in nonlinear medium is automatically eliminated, and only pure fundamental and second-order harmonics are maintained in the iteration, as shown in Fig.~\ref{fig:CFFD necessity}(c). This advantage of CFFD algorithm is very helpful for improving accuracy of nonlinear simulation.    

	In the above demonstration, the pseudo-frequency components in nonlinear calculations are triumphantly eliminated via introduction of complex field and frequency-decomposition procedure.

\subsection{Frequency-dependent nonlinearity}
\label{subsec:Frequency-dependent nonlinearity}
	When the fundamental field has a broad-band spectrum, according to nonlinear theory for perturbative and slowly-varying-envelope conditions, the sum-frequency wave satisfies the one-dimension wave equation\cite{Nonlinear_Optics_boyd}
	\begin{equation}
		\frac{dE_s(z)}{dz} = \frac{i\omega_s^2}{k_s c^2} \chi^{(2)}_{eff}(\omega_1,\omega_2) E_{\omega_1}(z) E_{\omega_2}(z) e^{-i\Delta k z},
	\end{equation}
	and the other two equations describing fundamental waves are negligible due to perturbative condition. Conveniently, we assume that this process is phase-matching, namely $\Delta k = 0$, and all three waves are without dispersion. Hence, the maximal amplitude of second harmonics varies along with propagating distance as 
	\begin{equation}\label{eq:Second-order nolinearity growth rate}
		\frac{dE_{s,M}(z)}{dz} = \frac{i\omega_s}{c\sqrt{\varepsilon_r}} \chi^{(2)}(\omega_1,\omega_2) E_{\omega_1,M} E_{\omega_2,M},
	\end{equation}
	where $E_{s,M}$, $E_{\omega_1,M}$ and $E_{\omega_2,M}$ represent maximal amplitude of $\omega_1 + \omega_2$, $\omega_1$ and $\omega_2$ frequency, respectively. Because the right-hand-side of Eq.~\eqref{eq:Second-order nolinearity growth rate} is a constant, $E_{s,M}$ increases linearly along with propagating distance $z$. We define analytical growth rate as 
	\begin{equation}\label{eq:growth rate expression}
		K_a(\omega_1,\omega_2) = \frac{\omega_s}{c\sqrt{\varepsilon_r(\omega_s)}} \chi^{(2)}_{eff}(\omega_1,\omega_2) E_{\omega_1,M} E_{\omega_2,M}.
	\end{equation}
	$K_a(\omega_1,\omega_2)$ is taken as the standard value for growth rate of second-order harmonics and compared with the numerical  $K_n(\omega_1,\omega_2)$ value obtained by averaging the growth values of different spatial points.

Here, we define the relative error $R = (K_n-K_a)/K_a$ to indicate the accuracy of CFFD algorithm. Based on this formula, we could utilize the perturbative CFFD and spatial filtering CFFD algorithms in sequence to simulate this nonlinear process for computing $R$. If $R$ is small enough, the validity of CFFD used for nonlinear calculation is confirmed.

\subsubsection{Perturbative CFFD}
\begin{figure} [ht]
		\begin{center}
		\hspace*{-0.7cm}
		\begin{tabular}{c}
			\includegraphics[scale=0.033]{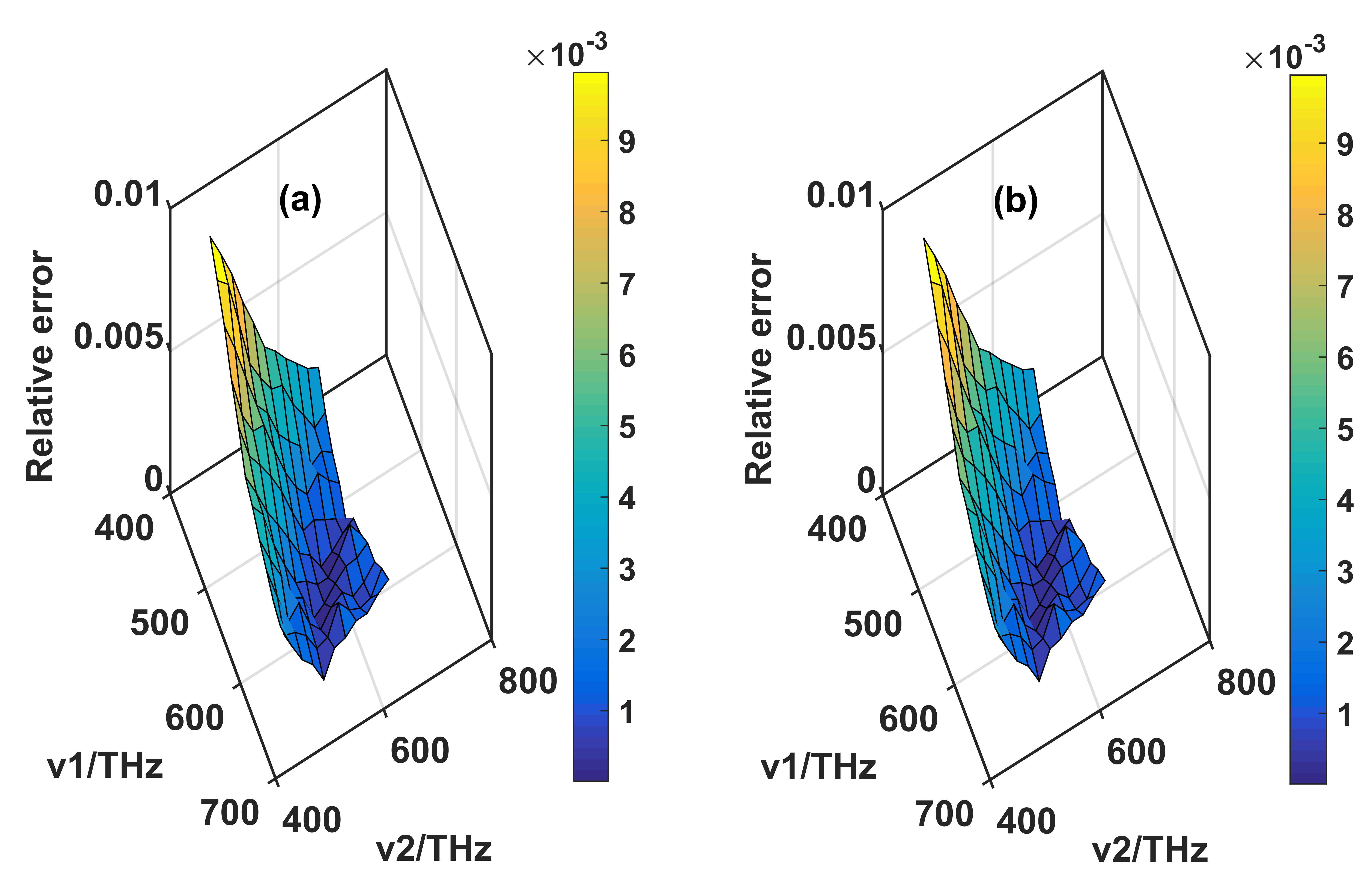} \\
			\includegraphics[scale=0.32]{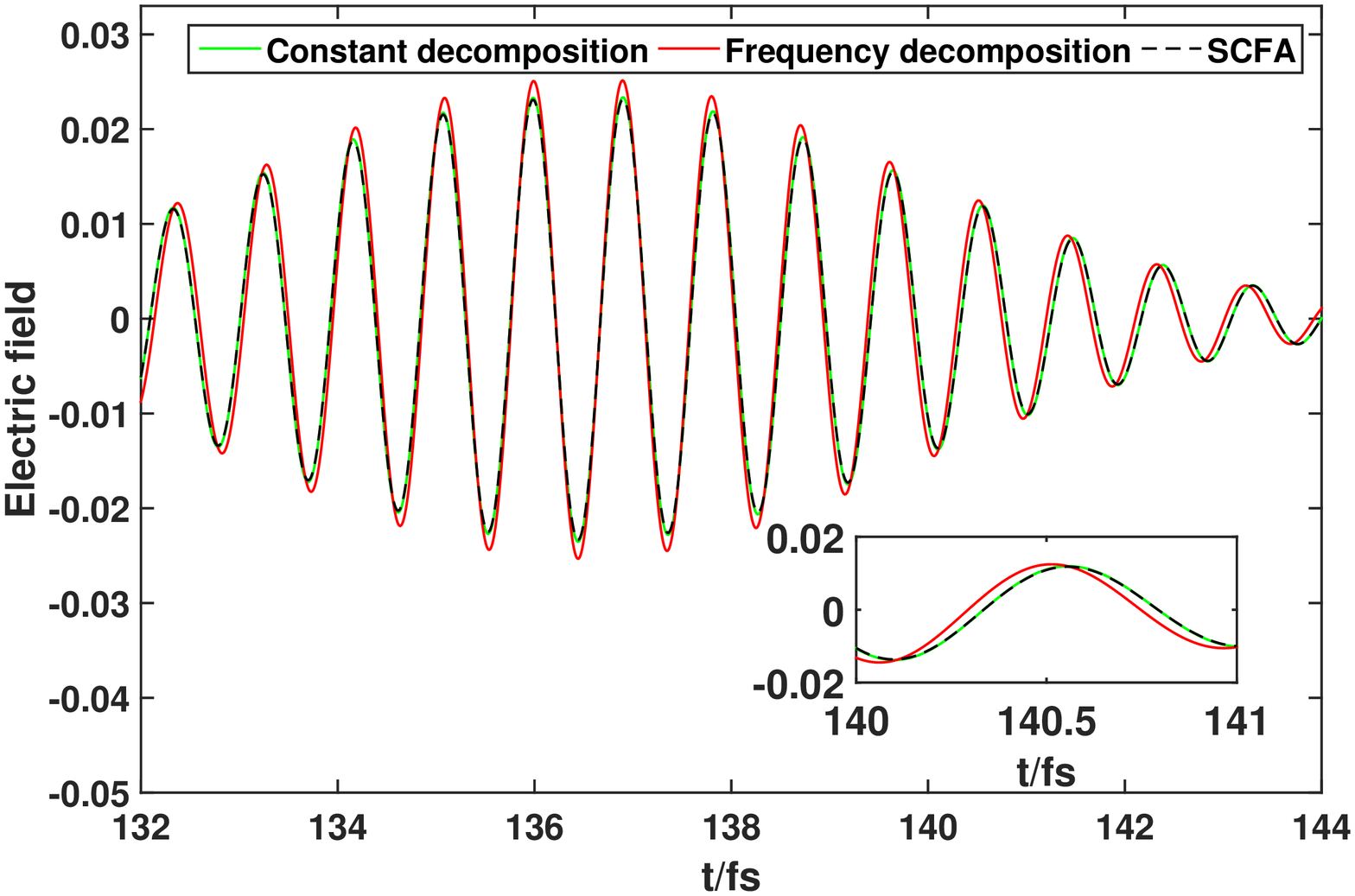}
		\end{tabular}
		\end{center}
		\caption{ \label{fig:perturbational CFFD accuracy} Results of Perturbative CFFD: (a) relative error distribution of growth rate for constant response; (b) relative error distribution of growth rate for frequency-dependent response; (c) temporal distribution for three algorithms (SCFA and constant decomposition are used for simulating constant response, frequency-dependent decomposition is used for frequency-dependent response).} 
	\end{figure}

	Now, we use the perturbative CFFD algorithm and the frequency-inde\-pendent nonlinear response is assumed to calculate the above nonlinear process. First, $f-\Delta F/2$ and $f+\Delta F/2$ are given as the left and right boundaries of frequency domain, and the whole calculating bandwidth $\Delta F = 200$~THz, sub-bandwidth $\Delta f = 20$~THz, the central frequency $f = 545$~THz, and the pulse duration $T_w = 7.6$~fs are adopted. Then, we choose different frequency pair $(\nu_1,\nu_2)$ to induce the corresponding sum-frequency component and then to compute the distribution of corresponding relative growth rate error $R$ as functions of $\nu_1$ and $\nu_2$, shown in Fig.~\ref{fig:perturbational CFFD accuracy}(a).

From this figure, one can see that the maximal error $R$ is around $8\times 10^{-3}$ appearing at boundary of  distribution by using the instantaneous response $\chi^{(2)} = A\times F(\omega_f)\times F(\omega_f)\times F(2\omega_f)$, where $F(\omega)$ is given by expression \eqref{eq:SecondNonlinearFormulation}. Really, the errors of boundary are indeed far less than the showed and approach to the central one, because the growth rate of expression \eqref{eq:growth rate expression} is an approximate formula for central frequency, which naturally lead to larger error outside the central area of $(\nu_f,\nu_f)$. Besides this, the error is negligible due to these parts with very small energy proportions. However, the relative error $R$ at the central frequency $(\nu_f,\nu_f)$, with large energy proportion, is only around $1\times10^{-4}$, which has negligible influence on the final result.

 According to this analysis, the final sum-frequency wave has a pretty accurate time-domain distribution (dotted line in Fig.~\ref{fig:perturbational CFFD accuracy}(c)) at $30$~$\mu$m away from input point, which is consistent with that obtained by the SCFA algorithm (dashed line). The SCFA algorithm has been confirmed that it has an accurate result for perturbative calculation \cite{SCFA_Ref1,SCFA_Ref2,SCFA_Ref3}. Hence, we can see again that CFFD algorithm has accuracy like what SCFA gives. 

However, if the frequency-dependent nonlinear response is considered, the SCFA algorithm would lose its ability, but the perturbative CFFD algorithm then shows its superior ability. Here, we provide the frequency-dependent nonlinear response as\cite{Nonlinear_Optics_boyd}
\begin{equation}\label{eq:SecondNonlinearFormulation}
	\begin{cases}
		\chi^{(2)}(\omega_1,\omega_2) = A\ F(\omega_1) F(\omega_2) F(\omega_1+\omega_2), \\
		F(\omega) = \dfrac{\omega_0^2}{\omega_0^2-\omega^2-2i\nu_0\omega}.
	\end{cases}
	\end{equation}
Moreover, we set $\nu_0=0$~THz, the dielectric intrinsic frequency $\omega_0 = 2\pi \times 1300$~THz, and response amplitude $A=1.6\times10^{-5}$~m/V. Following the same calculation procedure in Fig.~\ref{fig:perturbational CFFD accuracy}(a), the relative growth rate error $R$ as functions of $\nu_1$ and $\nu_2$, under frequency-dependent response, is shown in  Fig.~\ref{fig:perturbational CFFD accuracy}(b). It is found that the distribution of relative error $R$ is nearly identical with Fig.~\ref{fig:perturbational CFFD accuracy}(a), which indicates that the perturbative CFFD has similar accuracy when processing frequency-dependent and frequency-independent nonlinear responses. The corresponding time-domain distribution of the sum-frequency wave for frequency-dependent nonlinear response is also shown in Fig.~\ref{fig:perturbational CFFD accuracy}(c), which shows an obvious difference from those obtained in the former. It has a higher magnitude and slight compression of vibration frequency from both sides to the center, because there is a higher response value for the higher frequency than the lower frequency, as predicted by expression \eqref{eq:SecondNonlinearFormulation}.

These results sufficiently confirm that the perturbative CFFD algorithm has an evident advantage over the SCFA algorithm. However, if the sum-frequency wave has a too much broader bandwidth, continuing to use the perturbative CFFD algorithm is not suitable, because the response cannot be considered as constant. Moreover, if the nonlinearity is non-perturbative, the error will be amplified ceaselessly through iteration, because the superimposed frequency interval will become larger and larger with simulation going. Hence, the spatial filtering CFFD algorithm described in Section~\ref{subsec:Synchronously filtering method} should be adopted for those cases with broad-band and non-perturbative nonlinear response. 

\subsubsection{Spatial filtering CFFD}

\begin{figure} [ht]
		\begin{center}
		\hspace*{-0.7cm}
		\begin{tabular}{c}
			\includegraphics[scale=0.035]{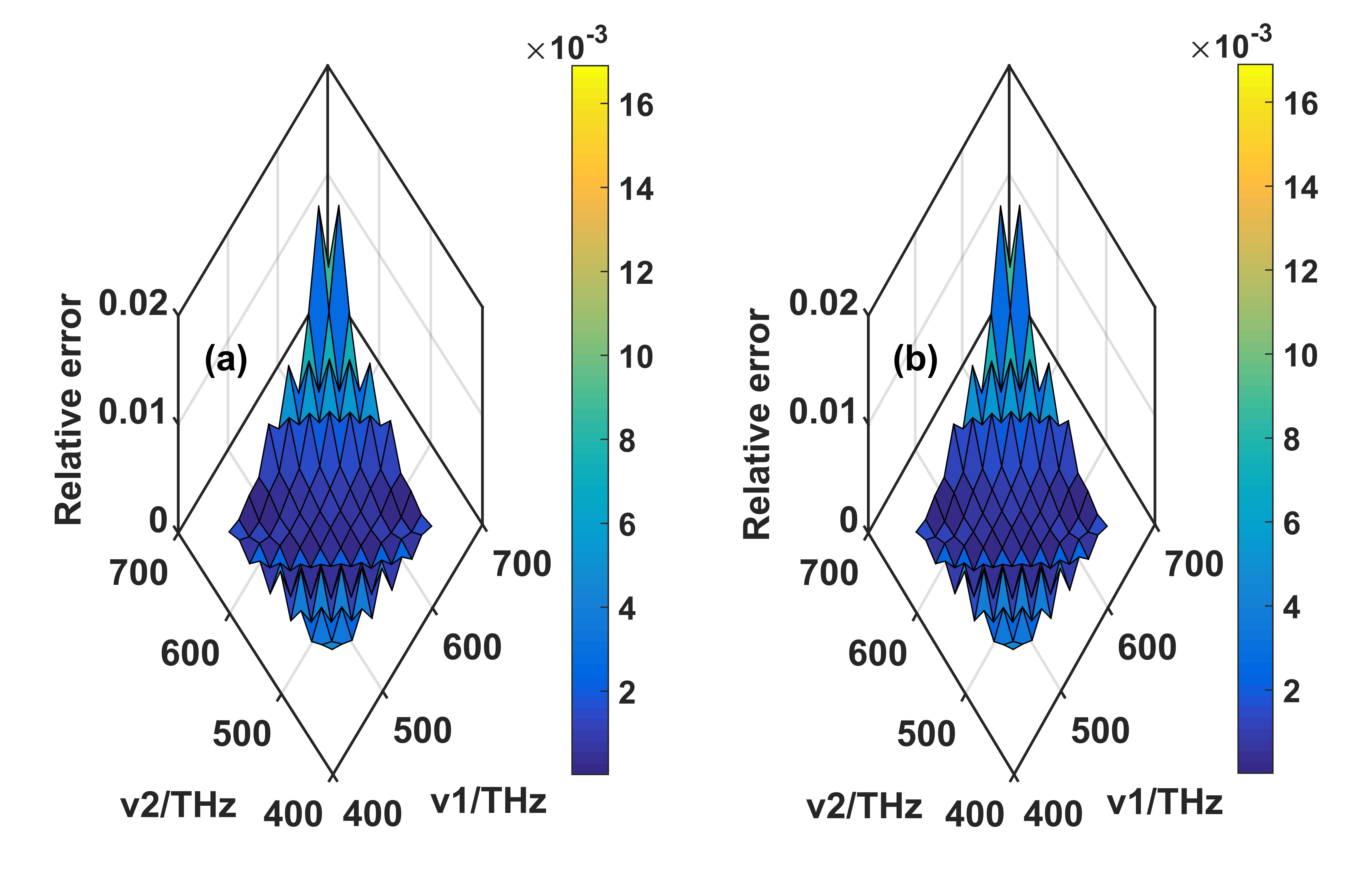} \\
			\includegraphics[scale=0.3]{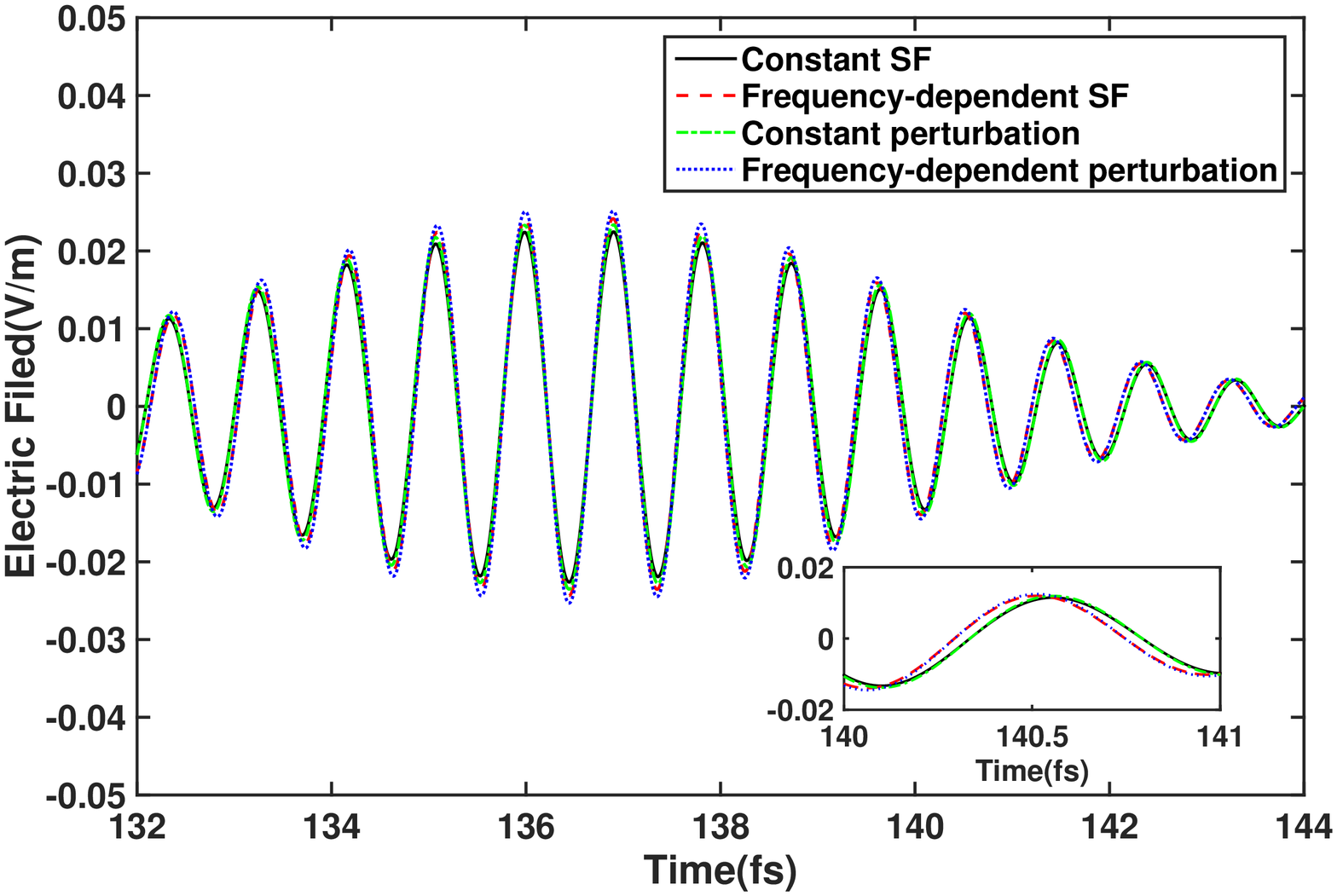}
		\end{tabular}
		\end{center}
		\caption{ \label{fig:Spatial Filtering CFFD accuracy} Results of spatial filtering CFFD: (a) relative error distribution of growth rate for constant response; (b) relative error distribution of growth rate for frequency-dependent response; (c) temporal distribution for four cases (the constant spatial filtering and the constant perturbative cases are used for constant response, the frequency-dependent spatial filtering and the frequency-dependent perturbative cases are used for frequency-dependent response).} 
	\end{figure}

	First, we use spatial filtering CFFD to compute second-order instantaneous response nonlinearity with $\chi^{(2)} = A\times F(\omega_f)\times F(\omega_f)\times F(2\omega_f)$. Then, in time domain the product of two electric fields with frequency intervals of $[\nu_1-\Delta f/2,\nu_1+\Delta f/2]$ and $[\nu_2-\Delta f/2,\nu_2+\Delta f/2]$ will produce a new sum-frequency electric field with frequency interval of $[\nu_1+\nu_2-\Delta f,\nu_1+\nu_2+\Delta f]$. Subsequently, we decompose it into left part with interval of $[\nu_1+\nu_2-\Delta f,\nu_1+\nu_2]$ and right part with interval of $[\nu_1+\nu_2,\nu_1+\nu_2+\Delta f]$ in time domain. Because the growth rate at each part cannot be computed directly, we  superpose the two electric fields in time domain corresponding to the respective left and right parts to get the maximum amplitude $E_{M}$ which are substituted into Eq.~\eqref{eq:growth rate expression} to get the growth rate $K_a$. The distribution of the relative error $R$ versus $\nu_1$ and $\nu_2$ is shown in Fig.~\ref{fig:Spatial Filtering CFFD accuracy}(a).

Like Fig.~\ref{fig:perturbational CFFD accuracy}(a), the relative error is also very small. For example, $R$ is only around $1\times 10^{-3}$ within the central part where the energy proportion is major.  At the boundary, the error is around $1\times 10^{-2}$, which is one order larger than that in the central part, which is because the analytical growth rate $K_a$ will deviate from the realistic value with frequency pair $(\nu_1,\nu_2)$ away from $(\nu_f,\nu_f)$, due to the breakdown of slowly-varying-envelope approximation. Despite this, it is unimportant because they have a very low energy proportion. Hence, we can say that the spatial filtering CFFD algorithm is very effective when processing band-broadening problems. In addition, the corresponding time-domain waveform obtained by constant spatial filtering CFFD (solid line in Fig.~\ref{fig:Spatial Filtering CFFD accuracy}(c)), is consistent with that by constant perturbative CFFD (dashed-dotted line).

	Second, we consider the frequency-dependent response with expression \eqref{eq:SecondNonlinearFormulation}. By following the same calculation procedure for Fig.~\ref{fig:Spatial Filtering CFFD accuracy}(a), the same parameters with Fig.~\ref{fig:perturbational CFFD accuracy}(b) are adopted to obtain the distribution of relative error $R$, which is shown in Fig.~\ref{fig:Spatial Filtering CFFD accuracy}(b). This distribution is almost same with Fig.~\ref{fig:Spatial Filtering CFFD accuracy}(a), which implies that this simulation is accurate like the Fig.~\ref{fig:Spatial Filtering CFFD accuracy}(a). The corresponding time-domain waveform (dashed line in Fig.~\ref{fig:Spatial Filtering CFFD accuracy}(c)) is compared with that by the perturbative CFFD algorithm (dotted line), and we find that they are consistent, which indicates the consistency of these two algorithms. 

From Fig.~\ref{fig:Spatial Filtering CFFD accuracy}(c), we can see that the waveform obtained by the spatial filtering CFFD algorithm has a slight difference from that by the perturbative CFFD algorithm, which is the direct consequence of the spatial filtering procedure. A little accuracy is sacrificed for the spatial filtering CFFD algorithm in order to enhance the ability of elimination of band-broadening effect and broadband response error. This kind of sacrifice is very necessary for the successful implementation of non-perturbative nonlinear calculation based on CFFD algorithm.

\subsection{Matched and mismatched phase conditions}
\label{subsec:Matched and mismatched phase conditions}
	Besides the analysis of relative error $R$, we can also prove the validity and advantage of CFFD algorithm through computing the coherent length $L_c$ for phase-mismatching condition.

\begin{figure} [ht]
		\begin{center}
		\begin{tabular}{c}
			\includegraphics[scale=0.5]{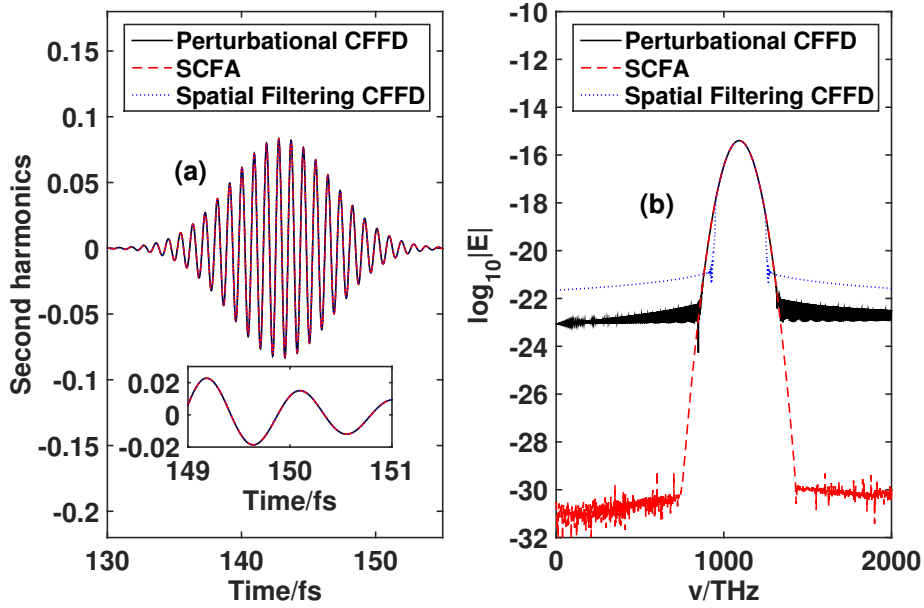}
		\end{tabular}
		\end{center}
		\caption{ \label{fig:SHG of different algorithm}Second harmonic simulation results by SCFA, Perturbative CFFD, and Spatial Filtering CFFD: (a) the temporal distributions are presented and show the consistency among them; (b) the spectral distributions of three methods show the strong consistency among them.}
	\end{figure}

	First, from temporal angle, a long Gaussian pulse with carrier frequency $\nu_f = 545$~THz and pulse width $T_w = 10$~fs is inputted to stimulate second-order harmonics. Then, the electric field at 30~$\mu$m away from the input point is recorded via three methods that the former adopted, including SCFA (dashed line), perturbative CFFD (solid line), and spatial filtering CFFD (dotted line), and is shown in Fig.~\ref{fig:SHG of different algorithm}(a), which sufficiently exhibits the significant consistency among them. In addition, the spectra of these electric fields are calculated and depicted in Fig.~\ref{fig:SHG of different algorithm}(b), which shows that the peak with central frequency $\nu_s = 1090$~THz is completely coincident within arrange from $\nu_l = 950$~THz to $\nu_r = 1250$~THz. There is a little of difference at the both sides of spectrum, which is because the truncated points of spatial filtering CFFD are selected at $\nu_l$ and $\nu_r$. If we change truncated points to enhance the width of interval $[\nu_l,\nu_r]$, a higher accuracy can be reached. Here, we reasonably select the arrange of filtering to maintain six magnitude-orders accuracy, which is accurate enough for applications due to the decomposed-fields characteristic that those fields with large contrast can be computed independently rather than calculated together.

\begin{figure} [ht]
		\begin{center}
		\hspace*{-1cm}
		\begin{tabular}{c}
			\includegraphics[scale=0.05]{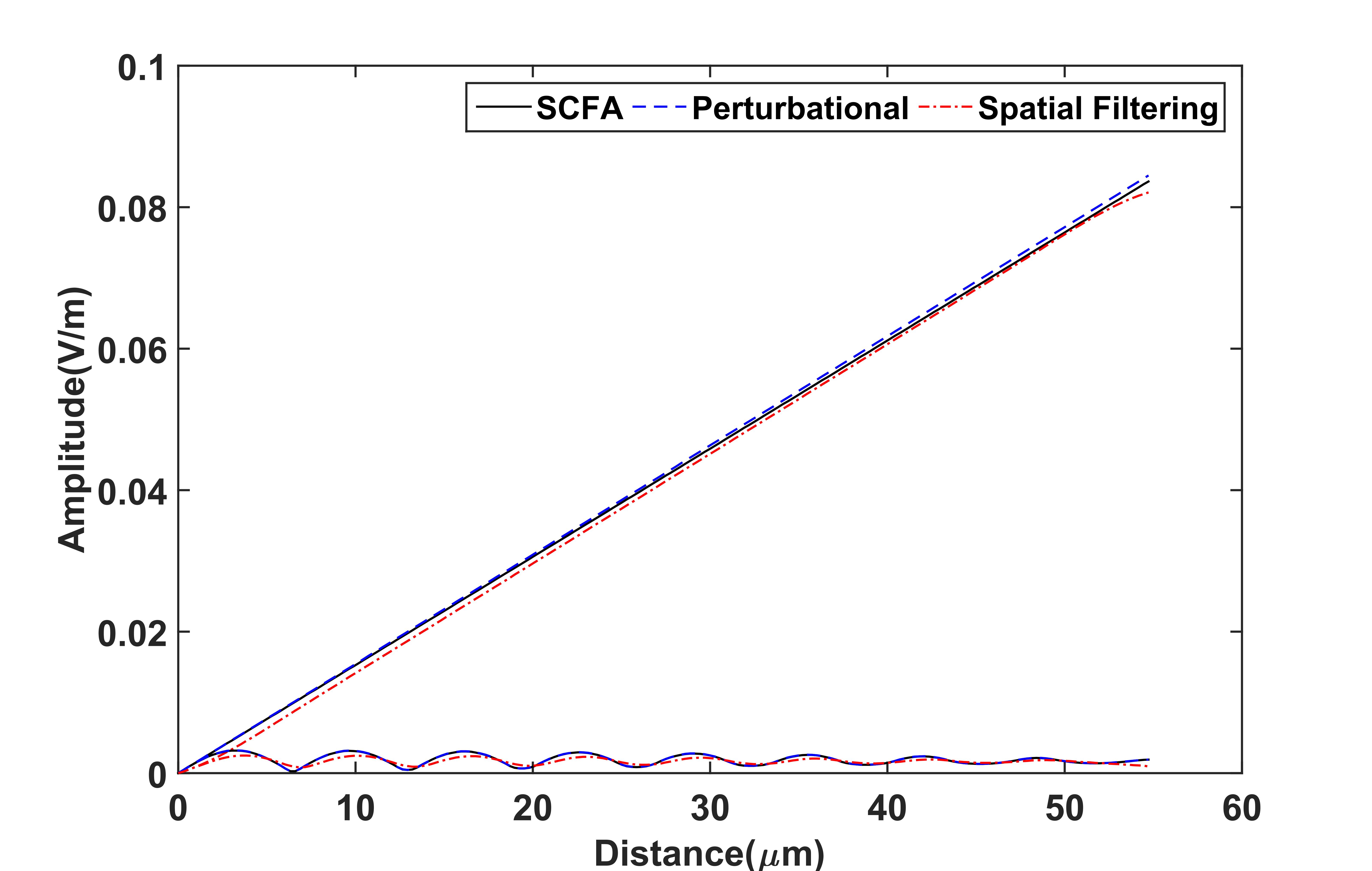}
		\end{tabular}
		\end{center}
		\caption{ \label{fig:SHG of different algorithm2}Second harmonic simulation results by SCFA, Perturbative CFFD, and Spatial Filtering CFFD. The straight lines are the growth rate of electric field amplitude under phase-matching condition; the wave lines present the varying periodicity of electric field amplitude under phase-mismatching conditions.}
	\end{figure}

	Second, from spatial angle, the correlation of maximum field amplitude $E_M(z)$ versus propagating distance $z$ is depicted in Fig.~\ref{fig:SHG of different algorithm2}. Under phase-matching case, with refractive indices $n_f = n_s = \sqrt{1.43}$, the electric field amplitudes by three methods are depicted as the three straight lines, which are consistent with what section~\ref{subsec:Frequency-dependent nonlinearity} gives. Another, under phase-mismatching case, with refractive indices $n_f = \sqrt{1.43}$ and $n_s = \sqrt{1.33}$, alternatively increasing and decreasing amplitudes are exhibited as the three wave lines in Fig.~\ref{fig:SHG of different algorithm2}, with propagating distance $z$ growing. We can see that the numerically computed periodicity is around $6.47~\mu m$ for spatial filtering CFFD, $6.43~\mu m$ for perturbative CFFD and $6.46~\mu m$ for SCFA. In addition, according to nonlinear theory, the coherent buildup length is determined by $L_c = \lambda_f/4(n_f-n_s)$, where $\lambda_f$ is wavelength of fundamental wave, $n_f$ and $n_s$ are refractive indices of fundamental and second harmonic waves. Hence, the theoretical periodicity is evaluated as $2L_c \approx 6.46~\mu m$, which is extremely close to the numerical values. Thus, we can say that CFFD algorithm is very accurate to simulate nonlinear processes for different phase-matching conditions, which proves its robustness again.

	\subsection{Summary}
	\label{subsec:Summary}
	In order to prove the validity of CFFD algorithm proposed in section~\ref{sec:theoretical background}, CFFD algorithm is first used to simulate a Lorentz dielectric and compared with SCFA algorithm and ADE algorithm in section~\ref{subsec:Lorentz response}, which sufficiently illustrates that CFFD algorithm is more consistent with ADE algorithm that has been widely applied in Lorentz linear dispersion, than SCFA algorithm. In section~\ref{subsec:Empirical response}, a Cauchy-like empirical medium that cannot be processed by conventional FDTD algorithm, like ADE algorithm, is simulated by CFFD algorithm, which satisfies to the theoretical results of reflected and transmitted coefficients predicted by Fresnel formula. The robustness of CFFD algorithm is presented sufficiently.

	Furthermore, CFFD algorithm has more evident advantages for computing nonlinear problems. Because of the simplicity of second-order nonlinearity, we use it to prove the accuracy and advantages of CFFD algorithm. In section~\ref{subsec:Elimination of pseudo-frequency components}, we present the second-order nonlinear results via using different methods and prove the advantage and necessity of complex-number field in canceling out pseudo-frequency components that seriously affect computing accuracy. In addition, CFFD algorithm not only can eliminate pseudo-frequency components, but also can process frequency-dependent nonlinear responses. In section~\ref{subsec:Frequency-dependent nonlinearity}, we use our CFFD algorithm to compute phase-matching second-order nonlinearity that cannot be accurately computed by traditional FDTD algorithm. Through presenting the relative error of growth rate between numerical and theoretical results, the advantage that CFFD algorithm can calculate frequency-dependent nonlinearity is sufficiently illuminated. In section~\ref{subsec:Matched and mismatched phase conditions}, we adopt CFFD algorithm to numerically compute the relation of electric field amplitude versus propagating distance under phase-matching and phase-mismatching conditions, respectively. Based upon this, the consistency between numerical and theoretical results for growth rate and coherent length is proved, which sufficiently demonstrates the robustness of CFFD algorithm for processing complicated nonlinearity.

\section{Conclusion}
\label{sec:conclusion}
	For resolving the difficulties of simulating ultra-wide-band linear dispersion and nonlinear response by FDTD, which is commonly faced by electromagnetic researchers, especially for those working in ultra-short lasers, this paper proposes an improved FDTD, CFFD-FDTD method. In section~\ref{sec:theoretical background}, we detailedly describe its theories, from the decomposition of Maxwell equations to polarization approximations for linear and nonlinear response, respectively. In section~\ref{sec:Applications of CFFD algorithm}, several characteristic examples are presented to illuminate the correctness and advantages of CFFD algorithm. For ultra-wide-band linear response, a trick of decomposition of the total spectrum into a series of sub-spectra that are with independent linear dispersion for each other is adopted, which is proved to be more robust than ADE algorithm. Subsequently, complex field is introduced to resolve the problem of pseudo-frequency components when processing nonlinearity. Furthermore, the difficulty of simulation of frequency-dependent nonlinearity is overcome by CFFD algorithm. Then, filtering CFFD is introduced in order to solve non-perturbative nonlinearity, especially spatial filtering method with very high accuracy, which is without non-physical divergence that conventional adiabatic approximation FDTD will produce.

	Overall, this new method not only has resolved the problems of calculation of ultra-wide-band linear dispersion and generation of pseudo-frequency components in nonlinearity, but also could simulate frequency-dependent nonlinearity, which is essential and important for simulating the interaction of ultra-short and ultra-strong laser with matters. Therefore, this algorithm provides an effective approach for those who are working in these or corresponding fields.

\section*{Acknowledgement}
\label{sec:acknowledgement}
	The work is supported by National Natural Science Foundation of China (Grant No. 12074398).

\bibliographystyle{elsarticle-num}
\nocite{*}
\bibliography{References}

\end{document}